\documentclass[12pt]{article}    % Specifies the document style.
\usepackage[ansinew]{inputenc} % permite utilizar los acentos y la ñ con el teclado directamente
\usepackage{hyperref}          % para poner enlaces web
\usepackage{graphics}
\usepackage{epsfig}
\usepackage{natbib}          % for citation processing
\bibpunct{(}{)}{;}{a}{,}{,}  % configuración de las citas
\usepackage{graphicx} %for graphics inclusion
\usepackage{amssymb}  %provides various useful mathematical symbols
\usepackage{amsmath}
\usepackage{multirow} %provides the command needed for spanning rows
\usepackage{enumerate}
\usepackage{color}     % para colorear el texto
\usepackage{indentfirst}  % para sangran el primer párrafo de cada sección
\usepackage{stackrel}          % (opc)  permite incluir fórmulas del tipo arg min
\begin{document}
%%%%%%%%%%%%%%%%%%%%%%%%%%%%%%%%%%%%%
\begin{center}

\section*{Modelling Road Accident Blackspots Data with the Discrete Generalized Pareto distribution}

\vskip 0.2in {\sc \bf Faustino Prieto$^a$
\footnote{Corresponding author. Tel.: +34 942 206758. Fax: +34 942 201603. E-mail address: faustino.prieto@unican.es (F. Prieto).},
Emilio G\'omez-D\'eniz$^b$, Jos\'e Mar\'{\i}a Sarabia$^a$

\vskip 0.2in

{\small\it $^a$Department of Economics, University of Cantabria, Avenida de los Castros s/n, 39005 Santander, Spain\\
           $^b$Department of Quantitative Methods in Economics, University of Las Palmas de Gran Canaria, 35017 Las Palmas de G.C., Spain.
}\\
}\end{center}

\begin{abstract}\noindent
This study shows how road traffic networks events, in particular road accidents on blackspots, can be modelled with simple probabilistic distributions.
We considered the number of accidents and the number of deaths on Spanish blackspots in the period 2003-2007, from Spanish General Directorate of Traffic (DGT).
We modelled those datasets, respectively, with the discrete generalized Pareto distribution (a discrete parametric model with three parameters)
and with the discrete Lomax distribution (a discrete parametric model with two parameters, and particular case of the previous model). For that, we analyzed the basic properties of both parametric models: cumulative distribution, survival, probability mass, quantile and hazard functions, genesis and rth-order moments; applied two estimation methods of their parameters: the $\mu$ and ($\mu+1$) frequency method and the maximum likelihood method; and used two goodness-of-fit tests: Chi-square test and discrete Kolmogorov-Smirnov test based on bootstrap resampling. We found that those probabilistic models can be useful to describe the road accident blackspots datasets analyzed.

\end{abstract}

\vskip 0.2in

\noindent {\bf Key Words}: Accident blackspots; Road traffic networks; Complex systems; Discrete Lomax distribution; Discrete generalized Pareto distributions.\\

\section{Introduction}

Road traffic networks are key economic drivers in today's world.
They provide a quick, reliable and flexible transportation system,
for people, goods and services. Unfortunately, traffic accidents
happen everyday - road injury was one of the top 10 causes of
death in the world in 2011, according to the World Health
Organization \citep{WHO}. For that reason, many research has been
devoted to the analysis of road accidents from different points of
views \citep[see, for
example,][and references therein]{AgueroValverde2013,Alemany2013,Brijs2007,Cafiso2011,Deublein2013,Kim2007,Li2012,Matirnez2013}.

In particular, the analysis of blackspots \citep[accident-prone
road sections,][]{Chen2011} has been considered  one of the basic
steps to reduce road accident rates. In this direction, several
methods to identify blackspots have been proposed \citep[accident
frequency method; accident rate methods; quality control method; empirical Bayesian method; and many
more;][]{Hauer2002,Cheng2005,Jurenoks2008,Sabel2005,Pei2005,Gregoriades2013};
Geographical Information Systems (GIS) have been incorporated in
the analysis of blackspots \citep{Chen2011,Mandloi2003}; the
effectiveness of blackspot programs has been evaluated in
different countries \citep{Meuleners2008,Hoque2007}; and it still
remains as an active field of research.

The aim of this study was to show that the road traffic networks
events, specifically the Spanish blackspots events, can be
described by simple probabilistic models. Although
the number of accidents and deaths on the road have decreased
considerably in Spain in the last years, due to the advertising
campaign promote by the authorities and the establishment of the
force of the points system on driving licences, which there has
been an improvement in responsible behaviour on the part of
drivers, the high rate of mortality in the Spanish roads
continues being very high. This fact motivates the study dealt
here, where we concentrate our attention in modelling: (1) the number of
accidents on Spanish blackspots, in the period 2003-2007, with the
discrete generalized Pareto distribution \citep[a discrete
parametric model with three parameters;][]{Asadi2001,Ekheden2012};
and (2) the number of deaths on Spanish
blackspots, in the same period 2003-2007, with the discrete Lomax
distribution \citep[a discrete parametric model with two
parameters, and particular case of the previous
model;][]{Krishna2009}. This paper also shows the basic
properties of both models; 
proposes two estimation methods of
their parameters: the $\mu$ and ($\mu+1$) frequency method and the
maximum likelihood method, leading the first one to initial
estimators for the second one; and recommends two
goodness-of-fit test for both of them: the Chi-square test and the
discrete Kolmogorov-Smirnov test.

The rest of this paper is organized as follows: in Section \ref{methods}, we introduce the dataset considered, the two probabilistic models used and the estimation methods and goodness of fit tests applied; finally, the results are presented and discussed in Section \ref{results}.

\section{Methods}\label{methods}

In this section,
we describe the road accident Spanish blackspots dataset used (section \ref{dat}),
we show the basic properties of the discrete generalized Pareto distribution (section \ref{dgp})
and the discrete Lomax distribution (section \ref{dlo}),
we propose two estimation methods of the parameters for both distributions considered (section \ref{est}),
and finally, we recommend two goodness-of-fit tests for those probabilistic distributions (section \ref{gof}).

\subsection{Road accident blackspot data}\label{dat}
We considered road accidents on Spanish blackspots data, from Spanish General Directorate of Traffic (D.G.T., 2003-2007) \citep{DGT1}, where blackspots are identificated by an accident frequency method as follows: a blackspot is a road section of 100 meters with three or more traffic accidents in a one-year period.

The dataset considered in this work consists of 16552 accidents, ocurred at Spanish road blackspots, from 2003 to 2007. In those accidents, unfortunately, 895 people died (within 30 days after accident) - which means, the 5.4 \% of the deaths in Spanish road accidents in that period \citep{DGT2}.

\begin{figure}[h]
\renewcommand{\figurename}{\footnotesize{Figure}}
\caption{\label{fig1}Number of Spanish road accident blackspots, and number of accidents and deaths on those blackspots in the period: 2003-2007.}
\centering
$\;$\\[1ex]
\includegraphics*[scale=0.4]{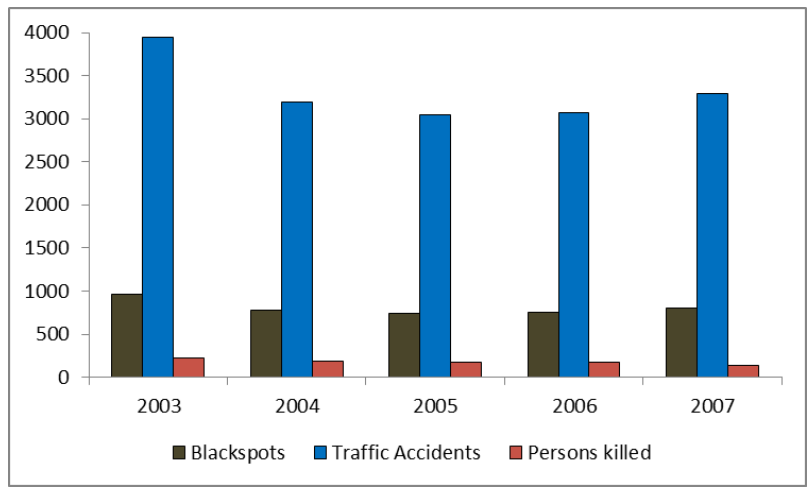}
\end{figure}

Table \ref{data1} and Fig.\ref{fig1} show the total number of Spanish blackspots, and the total number of accidents and deaths happened on those blackpots, in each year of the period considered \footnote{Note that we found four blackspots (one in the year 2003, two in the year 2005, and one more in the year 2006) with only two traffic accidents. in the dataset published by DGT. Due to the blackspot definition (3 or more accidents), we didn't include them in this study. However, we confirmed separately that the main conclusions reached in this work are the same including them as blackspots with three accidents instead of two.}

We analyzed two discrete variables. The first one was the number of road accidents on Spanish blackspots for the year considered - the corresponding data can be seen in Table \ref{data2} (for example, in 2003, there were 525 blackspots in Spain where 3 accidents ocurred). The second discrete variable was the number of deaths on Spanish blackspots for the year considered - see Table \ref{data3} (for example, in 2003, there were 797 blackspots where nobody died). Note that the first discrete variable selected (number of accidents) has the minimum value of 3 due to the blackspot definition, and the second discrete variable analyzed (number of deaths) has the minimum value of 0 corresponding to accidents without deaths.

\begin{table}[h]\tiny
  \renewcommand{\tablename}{\footnotesize{Table}}
  \renewcommand\arraystretch{1.5}
  \setlength{\tabcolsep}{4.0 mm}
  \caption{\label{data1}Number of Spanish road accident blackspots, and number of accidents and deaths on those blackspots in the period: 2003-2007.}
  \centering
        \begin{tabular*}{1.0\textwidth}{l c c c c c c}
        \\[-1ex]
     \hline
                                                   &   2003       & 2004        &   2005        &   2006       &   2007         &   Total   \\
     \hline
Number of blackspots in Spain                &   958        &   780       &   737         &   748        &   802          &                \\
Number of accidents on Spanish blackspots    &   3941       &   3200      &   3051        &   3071       &   3289         &      16552     \\
Number of deaths on Spanish blackspots       &    220       &    191      &    179        &    171       &    134         &        895     \\
     \hline
     \end{tabular*}
\end{table}

\begin{table}[p]\tiny
  \renewcommand{\tablename}{\footnotesize{Table}}
  \renewcommand\arraystretch{1.5}
  \setlength{\tabcolsep}{8.0 mm}
  \caption{\label{data2}Number of road accident blackspots in Spain according to the number of road accidents. Period: 2003-2007.}
  \centering
        \begin{tabular*}{1.0\textwidth}{c c c c c c}
        \\[-1ex]
     \hline
Number                                    &                   \multicolumn{5}{c}{Number of road accident blackspots in Spain}                        \\
  \cline{2-6}
of accidents                              &   2003               &    2004              &   2005               &   2006               &   2007        \\
     \hline
3                                         &   525                &   438                &   400                &   404                &   445          \\
4                                         &   209                &   173                &   177                &   164                &   172          \\
5                                         &   94                 &   71                 &   68                 &   89                 &   77           \\
6                                         &   41                 &   38                 &   35                 &   45                 &   48           \\
7                                         &   34                 &   23                 &   22                 &   20                 &   19           \\
8                                         &   15                 &   9                  &   11                 &   8                  &   11           \\
9                                         &   22                 &   8                  &   4                  &   4                  &   11           \\
10                                        &   5                  &   6                  &   4                  &   1                  &   2            \\
11                                        &   2                  &   1                  &   3                  &   3                  &   4            \\
12                                        &   4                  &   3                  &   3                  &   2                  &   1            \\
13                                        &   1                  &   2                  &   3                  &   --                 &   5            \\
14                                        &   1                  &   2                  &   --                 &   1                  &   2            \\
15                                        &   --                 &   2                  &   1                  &   --                 &   2            \\
16                                        &   1                  &   --                 &   2                  &   2                  &   --           \\
17                                        &   1                  &   --                 &   --                 &   --                 &   1            \\
18                                        &   --                 &   --                 &   1                  &   --                 &   --           \\
19                                        &   1                  &   1                  &   --                 &   --                 &   --           \\
20                                        &   1                  &   1                  &   --                 &   --                 &   --           \\
21                                        &   --                 &   --                 &   --                 &   1                  &   --           \\
22                                        &   --                 &   --                 &   --                 &   1                  &   --           \\
24                                        &   --                 &   --                 &   --                 &   1                  &   --           \\
25                                        &   --                 &   --                 &   --                 &   --                 &   1            \\
27                                        &   --                 &   1                  &   --                 &   --                 &   --           \\
29                                        &   --                 &   --                 &   --                 &   1                  &   --           \\
32                                        &   --                 &   --                 &   --                 &   --                 &   1            \\
33                                        &   --                 &   --                 &   2                  &   --                 &   --           \\
36                                        &   --                 &   --                 &   1                  &   --                 &   --           \\
39                                        &   1                  &   --                 &   --                 &   1                  &   --           \\
49                                        &   --                 &   1                  &   --                 &   --                 &   --           \\[1ex]
Total                                     &   958                &   780                &   737                &   748                &   802          \\
    \hline
     \end{tabular*}
\end{table}

\begin{table}[p]\tiny
  \renewcommand{\tablename}{\footnotesize{Table}}
  \renewcommand\arraystretch{1.5}
  \setlength{\tabcolsep}{8.2 mm}
    \caption{\label{data3}Number of road accident blackspots in Spain according to the number of deaths. Period: 2003-2007.}
  \centering
        \begin{tabular*}{1.0\textwidth}{c c c c c c}
        \\[-1ex]
     \hline
Number                                    &                   \multicolumn{5}{c}{Number of road accident blackspots in Spain}                        \\
  \cline{2-6}
of deaths                                &   2003               &    2004              &   2005               &   2006               &   2007        \\
  \hline
0&  797&    636&    611&    632&    693\\
1&  126&    108&     94&     84&   92\\
2&   19&   27&   23&     19&     12\\
3&   12&      7&      3&      9&      4\\
4&    2&      2&      3&      1&     --\\
5&   --&     --&      1&      1&     --\\
6&    2&     --&      1&      1&      1\\
7&   --&     --&      1&      1&     --\\[1ex]
Total                                     &   958                &   780                &   737                &   748                &   802          \\
     \hline
            \end{tabular*}
\end{table}

\subsection{The discrete generalized Pareto distribution}\label{dgp}

The discrete generalized Pareto distribution is defined in terms of the cumulative distribution function (cdf). If $F(x)=\Pr(X\leq x)$ we define:
\begin{equation}\label{cdf}
F(x)=1-[1+\lambda(x-\mu+1)]^{-\alpha},\;\;x=\mu,\mu+1,\dots
\end{equation}
and $F(x)=0$ if $x<\mu$, where $\alpha,\lambda,\mu>0$ are respectively the shape, scale and location parameters.

A random variable with cdf (\ref{cdf}) will be denoted by $X\sim
{\cal DGP}(\alpha,\lambda,\mu)$.

The survival function, $\bar F(x)=\Pr(X\geq x)=1-F(x-1)$\citep{Xekalaki1983}, of the ${\cal DGP}$ distribution can be written as
\begin{equation}\label{sf}
\bar F(x)=[1+\lambda(x-\mu)]^{-\alpha},\;\;x=\mu,\mu+1,\dots
\end{equation}

The probability mass function (pmf) of the ${\cal DGP}$ distribution is given by
\begin{equation}\label{pmf}
\Pr(X=x)=[1+\lambda(x-\mu)]^{-\alpha}-[1+\lambda(x-\mu+1)]^{-\alpha},\;\;x=\mu,\mu+1,\dots
\end{equation}

The ${\cal DGP}$ distribution is unimodal with a modal value at $x=\mu$. Letting Eq.(\ref{pmf}), define
$\Pr(X=x)$ also for non-integer values of $x$, then we obtain that for $x\geq \mu$
$$\displaystyle\frac{d\Pr(X=x)}{dx}=\alpha\lambda[(1+\lambda(x-\mu+1))^{-(\alpha+1)}-(1+\lambda(x-\mu))^{-(\alpha+1)}],$$
which clearly is negative, and hence the probability mass function is a decreasing function on $X$.

\begin{figure}[p]
\renewcommand{\figurename}{\footnotesize{Figure}}
\caption{\label{fig2}Probability mass function of the ${\cal DGP}$ distribution for selected values of
$\mu,\alpha,\lambda$.} \centering
\includegraphics*[scale=1.0]{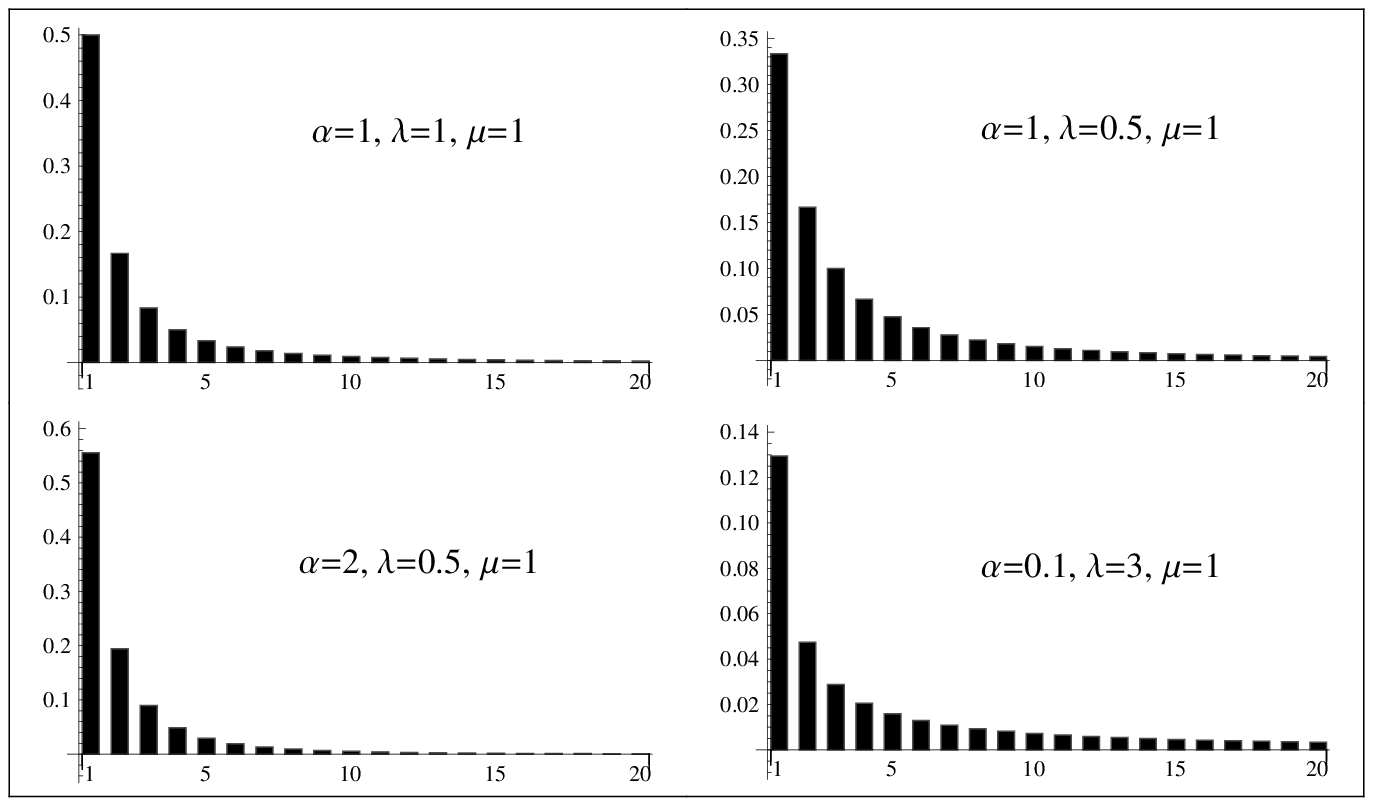}
\end{figure}

Fig.\ref{fig2} shows the different forms of pmf given by
Eq.(\ref{pmf}), according to different values of the parameters
$\alpha$ and $\lambda$.

The ${\cal DGP}$ distribution can be obtained by
discretizing the continuous generalized Pareto
distribution, by applying the following expression
\citep{Kemp2008,Nakagawa1975,GomezDeniz2010,GomezDeniz2011b,GomezDeniz2011,Roy2004}

\begin{equation}\label{genesis}
\Pr(X=x)=\bar{F}(x)-\bar{F}(x+1),
\end{equation}
where $\bar{F}(x)$ is the survival function of that continuous model and $\Pr(X=x)$ is the probability mass function of the discrete model associated. Using Eq.(\ref{genesis}) with the survival function of the continuous generalized Pareto distribution \citep{Arnold1983}, given by $\bar{F}(x)=[1+\lambda(x-\mu)]^{-\alpha}$, we obtain the pmf of the ${\cal DGP}$ distribution given in the Eq.(\ref{pmf}).

The quantile function can be obtained by inverting Eq.(\ref{cdf}),
\begin{equation}\label{guantil}
x_\gamma=\left\lceil\displaystyle\frac{(1-\gamma)^{-1/\alpha}-1}{\lambda}-1+\mu\right\rceil,
\end{equation}
where $\left\lceil\;\right\rceil$ denotes the ceiling function (the smallest integer greater than or equal). In particular, the median is given by
$$x_{0.5}=\left\lceil\displaystyle\frac{2^{1/\alpha}-1}{\lambda}-1+\mu\right\rceil.$$

The hazard function (failure rate), using Eqs.(\ref{sf}) and
(\ref{pmf}), can be written as
\begin{eqnarray}r(x)=\displaystyle\frac{Pr(X=x)}{\bar
F(x)}=1-\left[\displaystyle\frac{1+\lambda(x-\mu)}{1+\lambda(x-\mu+1)}\right]^{\alpha},\;\;x=\mu,\mu+1,\dots\label{hrf}
\end{eqnarray}
which is a decreasing function on $X$.

The rth-order moment of ${\cal DGP}$ distribution
becomes
\begin{eqnarray}
E(X^r) &=&
\sum^{\infty}_{x=\mu}x^r\Pr(X=x)=\sum^{\infty}_{x=\mu+1}[x^r-(x-1)^r]\bar{F}(x)\nonumber\\
&=& \sum^{\infty}_{x=\mu+1}\frac{x^r-(x-1)^r}
{[1+\lambda(x-\mu)]^{\alpha}}. \label{rmoment}
\end{eqnarray}

In particular, the mean and the 2nd moment are given by
$$E(X)=\sum^{\infty}_{x=\mu+1}\frac{1}{[1+\lambda(x-\mu)]^{\alpha}},$$
$$E(X^2)=\sum^{\infty}_{x=\mu+1}\frac{2x-1}{[1+\lambda(x-\mu)]^{\alpha}},$$
which are finite, respectively, if $\alpha>1$ and $\alpha>2$. Note that $\partial E(X)/\partial \alpha<0$ and $\partial E(X)/\partial \lambda<0$, and therefore the mean decreases with both parameter: $\alpha$ and $\lambda$.

Table \ref{indexdispersion} shows the index of dispersion $D=[E(X^2)-(E(X))^2]/E(X)$, for different values of the parameters $\alpha$ and $\lambda$. It can be seen that this variance-to-mean ratio seems always be larger than 1, and therefore the ${\cal DGP}$ distribution seems overdispersed.
\begin{table}[p]\tiny
  \renewcommand{\tablename}{\footnotesize{Table}}
  \renewcommand\arraystretch{1.5}
  \setlength{\tabcolsep}{5.0 mm}
  \caption{\label{indexdispersion}Index of dispersion $D=[E(X^2)-(E(X))^2]/E(X)$, for different values of $\alpha$,$\lambda$.}
  \centering
        \begin{tabular*}{1.0\textwidth}{c c c c c c c c c}
        \\[-1ex]
     \hline
                                  &                   \multicolumn{8}{c}{{\bf \scriptsize $\alpha$}}                                                \\
     \cline{2-9}
{\bf \scriptsize $\lambda$}            &{\bf 3}    &{\bf 4}    &{\bf 5}    &{\bf 6}    &{\bf 7}    &{\bf 8}    &{\bf 9}    &{\bf 10}    \\
     \hline
{\bf 0.1}&16.47& 7.70& 5.04& 3.79& 3.08& 2.63& 2.31& 2.08\\
{\bf 0.2}& 9.05& 4.40& 2.99& 2.33& 1.96& 1.72& 1.56& 1.44\\
{\bf 0.3}& 6.58& 3.31& 2.32& 1.86& 1.60& 1.44& 1.33& 1.25\\
{\bf 0.4}& 5.36& 2.77& 1.99& 1.63& 1.43& 1.30& 1.22& 1.16\\
{\bf 0.5}& 4.63& 2.45& 1.80& 1.49& 1.33& 1.23& 1.16& 1.11\\
{\bf 0.6}& 4.14& 2.24& 1.67& 1.41& 1.26& 1.18& 1.12& 1.08\\
{\bf 0.7}& 3.79& 2.09& 1.58& 1.34& 1.22& 1.14& 1.09& 1.06\\
{\bf 0.8}& 3.54& 1.98& 1.51& 1.30& 1.19& 1.12& 1.08& 1.05\\
{\bf 0.9}& 3.34& 1.89& 1.46& 1.26& 1.16& 1.10& 1.06& 1.04\\
{\bf 1.0}& 3.18& 1.82& 1.42& 1.24& 1.14& 1.09& 1.05& 1.03\\
{\bf 2.0}& 2.45& 1.51& 1.24& 1.12& 1.06& 1.03& 1.02& 1.01\\
{\bf 3.0}& 2.21& 1.41& 1.18& 1.09& 1.04& 1.02& 1.01& 1.00\\
{\bf 4.0}& 2.09& 1.36& 1.15& 1.07& 1.03& 1.02& 1.01& 1.00\\
{\bf 5.0}& 2.02& 1.33& 1.14& 1.06& 1.03& 1.01& 1.00& 1.00\\
{\bf 6.0}& 1.97& 1.31& 1.13& 1.06& 1.03& 1.01& 1.00& 1.00\\
{\bf 7.0}& 1.94& 1.29& 1.12& 1.05& 1.02& 1.01& 1.00& 1.00\\
{\bf 8.0}& 1.91& 1.28& 1.12& 1.05& 1.02& 1.01& 1.00& 1.00\\
{\bf 9.0}& 1.89& 1.28& 1.11& 1.05& 1.02& 1.01& 1.00& 1.00\\
{\bf10.0}& 1.87& 1.27& 1.11& 1.05& 1.02& 1.01& 1.00& 1.00\\
     \hline
     \end{tabular*}
\end{table}

\subsection{The discrete Lomax distribution}\label{dlo}

The class defined in Eq.(\ref{cdf}) includes an important 2-parameter particular case: the discrete Lomax distribution ($\mu=0$).

The discrete Lomax distribution is also defined in terms of the cumulative distribution function (cdf). If $F(x)=\Pr(X\leq x)$ we define:
\begin{equation}\label{cdf2}
F(x)=1-[1+\lambda(x+1)]^{-\alpha},\;\;x=0,1,\dots
\end{equation}
and $F(x)=0$ if $x<0$, where $\alpha,\lambda>0$ are respectively
the shape and scale parameters.

A random variable with cdf
(\ref{cdf2}) will be denoted by $X\sim {\cal DL}o(\alpha,\lambda)$. In this case ${\cal DGP}(\alpha,\lambda,0)\equiv{\cal DL}o(\alpha,\lambda)$.

The survival function is given by 
$$\bar F(x)=\Pr(X\geq x)=[1+\lambda x]^{-\alpha},\;\;x=0,1,\dots$$ 
The pmf of the ${\cal DL}o$ distribution is given by
\begin{eqnarray}
\Pr(X=x)=[1+\lambda
x]^{-\alpha}-[1+\lambda(x+1)]^{-\alpha},\;\;x=0,1,\dots
\label{dld}
\end{eqnarray}

The ${\cal DL}o$ distribution can be obtained also by discretizing the continuous Lomax
distribution (also known as Pareto II distribution,\citep{Arnold1983}), by applying the Eq.(\ref{genesis}). 

Of course that because the ${\cal DL}o$ is a particular case of the ${\cal DGP}$ distribution, 
again the distribution with unimodal
with a zero vertex and the quantile, hazard function and the
moments are obtained by putting $\mu=0$ in Eqs.(\ref{guantil}),
(\ref{hrf}) and (\ref{rmoment}), respectively.

Furthermore, a more simple distribution can be obtained directly
from Eq.(\ref{dld}) by assuming that $\alpha=1$ and obtained
therefore a one-parameter distribution which has good properties.
Although the mean of this distribution does not exist we will see
some interesting properties of this distribution can be obtained
easily. To the best of our knowledge, the discrete distribution
presented here has not been previously studied in detail in statistical
literature.

In this case, after straightforward computation we get the
probability mass function of this simple distribution given by
\begin{eqnarray}
\Pr(X=x)=\frac{\lambda}{(1+\lambda x)(1+\lambda(x+1))},\quad
x=0,1,\dots;\;\lambda>0. \label{ndd1}
\end{eqnarray}

The quantiles, hazard function and the median are obtained from
Eqs.(\ref{guantil}), (\ref{hrf}) and (\ref{rmoment}), respectively,
after putting $\alpha=1$ and $\mu=0$. In this particular case,
some algebra provides the probability generating function, which
is given by
\begin{equation*}
G_X(z)=\frac{1}{z}\left[1-(1-z)\Phi(z,1,1/\lambda)\right],\quad
|z|<1,\label{pgf}
\end{equation*}
where $\Phi(z,s,a)$ is the Hurwitz–Lerch transcendent function
given by $ \Phi(z,s,a)=\sum_{k=0}^{\infty}z^k(k+a)^{-s}.$

Because non-closed form exists for the moments of the
distribution, alternatively we can use the following inverse moment
\begin{equation*}
E\left(\frac{1}{X+1}\right)=\frac{\lambda  \left(\lambda +\psi
\left(1/\lambda\right)+\gamma -1\right)}{1-\lambda},
\end{equation*}
where $\psi(z)=\Gamma^{\prime}(z)/\Gamma(z)$ is the digamma
function and $\gamma$ is Euler's constant, with approximate
numerical value 0.577216. It must be reported that the
Hurwitz-Lerch transcendent function and the digamma function are
available in the Mathematica package. Additionally, since
$\Pr(X=0)\neq0,\;\Pr(X=1)\neq0$
and $\Pr(X=j)/\Pr(X=j-1),\;j=1,2,\dots$ forms a monotone increasing sequence,
we have that the distribution is infinitely divisible (log-convex). See \citep{WardeandKatti1971} for details. 
Therefore, $\Pr(X=x)$ is a decreasing sequence (see \citep{JohnsonandKotz1982}, p.75), which
is congruent with the zero vertex of the new distribution.
Moreover, as any infinitely divisible distribution defined on
nonnegative integers is a compound Poisson distribution (see
Proposition 9 in \citep{KarlisandXekalaki2005}), we conclude that
the new probability mass function (pmf) given in Eq.(\ref{ndd1}) is a compound Poisson
distribution.

Furthermore, the infinitely divisible distribution plays an
important role in many areas of statistics, for example, in
stochastic processes and in actuarial statistics. When a
distribution $G$ is infinitely divisible then for any integer
$x\geq 2$, there exists a distribution $G_x$ such that $G$ is the
$x$-fold convolution of $G_x$, namely, $G = G_x^{\ast x}$.

Finally, since the new distribution is infinitely divisible, a
lower bound for the variance can be obtained when $\alpha=1$ (see
\citep{JohnsonandKotz1982}, p.75), which is given by 
$var(X)\geq \frac{\Pr(X=1)}{\Pr(X=0)}=\frac{1}{1+\lambda}.$

\subsection{Estimation of parameters}\label{est}

Let $x_1,\dots,x_n$ be a sample of size $n$ drawn from a ${\cal DGP}$ distribution and $x_{min}$ the smallest
observation ($x_i\geq x_{min},\forall i$). We assume that $\mu$-parameter is given and can be estimated using the sample
minimum $\hat{\mu}=x_{min}$ (in the case of ${\cal DL}o$ distribution, $\hat{\mu}=x_{min}=0$). In consequence, we estimate both
parameters $\alpha$ and $\lambda$. For that, we propose two estimation methods: the $\mu$-frequency and ($\mu+1$)-frequency
method; and the maximum likelihood method. The first method leads to simple estimators, which can be used as initial estimators, the
seed value ($\alpha_0,\lambda_0$), in the maximum likelihood method.

To apply the $\mu$-frequency and ($\mu+1$)-frequency method for estimating the parameters $\alpha$ and $\lambda$, we have to calculate the relative frequencies of $X=\mu$ and of $X=(\mu+1)$ in the sample, which we denote as $\hat{p}_{\mu}$ and $\hat{p}_{\mu+1}$ respectively, equate them to the corresponding pmf values from Eq.(\ref{pmf}), and solve the two equations obtained, simultaneously for $\alpha$ and $\lambda$. From Eq.(\ref{pmf}):
\begin{eqnarray}
\hat{p}_{\mu}&=&1-[1+\lambda)]^{-\alpha},\label{seed1a}\\[1ex]
\hat{p}_{\mu+1}&=&[1+\lambda)]^{-\alpha}-[1+2\lambda)]^{-\alpha}.\label{seed1b}
\end{eqnarray}
If we eliminate $\alpha$ in Eqs.(\ref{seed1a}) and (\ref{seed1b}), we obtain the equation in $\lambda$,
\begin{equation}\label{seed1c}
\displaystyle\frac{\log(1+2\lambda)}{\log(1+\lambda)}=\displaystyle\frac{1-\hat{p}_{\mu}-\hat{p}_{\mu+1}}{1-\hat{p}_{\mu}}.
\end{equation}
The previous equation can be solved by a simple computer algorithm, taking in account that the left hand of the equation(\ref{seed1c}) is a monotone function in $\lambda$. Finally, the $\alpha$ estimator is (using Eq.(\ref{seed1a})):
\begin{equation}\label{seed1d}
\hat{\alpha}=-\displaystyle\frac{\log(1-\hat{p}_{\mu})}{\log(1+\hat{\lambda})}.
\end{equation}
Now, we consider the second method proposed: the maximum likelihood method. The log-likelihood function is given by,
\begin{eqnarray*}\label{mle}
\log\ell(\lambda,\nu)&=&\sum_{i=1}^{n}\log \Pr(X=x_i)=\\[-1ex]
                     &=& \sum^{n}_{1=1}\log\left[(1+\lambda(x_i-\mu))^{-\alpha}-(1+\lambda(x_i-\mu+1))^{-\alpha}\right].
\end{eqnarray*}
where $\Pr(X=x_i)$ is the pmf defined in Eq.(\ref{pmf}).

Taking partial derivatives with respect to $\alpha$ and $\lambda$, and equating them to zero, we obtain
the normal equations:
\begin{eqnarray*}
\frac{\partial\log\ell}{\partial\alpha}
&=&\sum_{i=1}^{n}\displaystyle\frac{\log\left[1+\lambda(x_i-\mu+1)\right]}{\left[1+\lambda(x_i-\mu+1)\right]^{\alpha}\left[1+\lambda(x_i-\mu)\right]^{-\alpha}-1}\\
&-&\sum_{i=1}^{n}\displaystyle\frac{\log\left[1+\lambda(x_i-\mu)\right]}{1-\left[1+\lambda(x_i-\mu)\right]^{\alpha}\left[1+\lambda(x_i-\mu+1)\right]^{-\alpha}}=0,\\
\frac{\partial\log\ell}{\partial\lambda}
&=&\sum_{i=1}^{n}\displaystyle\frac{\alpha(x_i-\mu+1)}{\left[1+\lambda(x_i-\mu+1)\right]^{\alpha+1}\left[1+\lambda(x_i-\mu)\right]^{-\alpha}-\left[1+\lambda(x_i-\mu+1)\right]}\\
&-&\sum_{i=1}^{n}\displaystyle\frac{\alpha(x_i-\mu)}{\left[1+\lambda(x_i-\mu)\right]-\left[1+\lambda(x_i-\mu)\right]^{\alpha+1}\left[1+\lambda(x_i-\mu+1)\right]^{-\alpha}}=0,
\end{eqnarray*}
which can be solved to obtain the maximum likelihood estimators.

In this study, maximum likelihood estimates of $\alpha$ and $\lambda$ were computed by numerical methods, using the \citep{Mathematica2010} software function \emph{FindMaximum}; and taking, as the initial value ($\alpha_0,\lambda_0$), the seed value obtained from the Eqs.(\ref{seed1c}) and (\ref{seed1d}) by the $\mu$-frequency and ($\mu+1$)-frequency method.

\subsection{Goodness-of fit test for ${\cal DGP}$ distribution}\label{gof}

Let $x_1,\dots,x_n$ be a sample of size $n$ of the discrete random variable $X$. In the previous section, we have assumed that the data come from the discrete generalized Pareto distribution (or from the discrete Lomax distribution) and we have shown how to estimate the parameters $\alpha,\lambda,\mu$ of that distribution. Therefore, a crucial task is to test the goodness of fit (GOF) of the sample with the ${\cal DGP}$ (or ${\cal DL}o$) model. For that, in this study we have used two different GOF tests: the Chi-Square test and the discrete Kolmogorov-Smirnov test.

The Chi-Square GOF test was first proposed by Karl Pearson in 1990 \citep{Pearson1900}. It gives us a measure of how close the observed values are to the expected values given by the fitted discrete model.\\
The chi-square GOF test statistics is as follows
\begin{equation*}
\chi^2=\sum_{i=1}^{k}\displaystyle\frac{(O_i-E_i)^2}{E_i},
\end{equation*}
where $O_i$ and $E_i$ are, respectively, the observed and the expected frequency for bin $i$ (with bins combined if the expected frequency is less than 5, and $k$ is the resulting number of bins). Note that the expected frequency was calculated, in this study, with the maximum likelihood estimator obtained in the previous section.\\
The null hypothesis can be expressed as
$$H_0: the\;data\;follow\;the\;discrete\;{\cal DGP}\;(or\; {\cal DL}o)\;model,$$
where null hypothesis can be rejected with the selected level of significance $\alpha=0.05$ if
\begin{equation*}
\chi^2>\chi_{0.95,k-r-1}^{2}\\
\end{equation*}
where $r$ is the number of parameter of the model fitted ($r=3$ for ${\cal DGP}$ distribution and $r=2$ for ${\cal DL}o$ distribution), and $\chi_{0.95,k-r-1}^{2}$ is the chi-square critical value with $k-r-1$ degrees of freedom.\\
Additionally, the corresponding $p$-value can be obtained, and null hypothesis can be rejected with the selected level of significance if $p$-value$<0.05$.\\

The discrete Kolmogorov-Smirnov ($KS$) test was first presented, as fas far as we know, by Norbert Henze in 1996 \citep{Henze1996}.
It is an empirical distribution function (EDF) goodness-of-fit (GOF) test for discrete data, based on the use of a parametric bootstrap.\\
The discrete $KS$ GOF test statistics for ${\cal DGP}$ (or ${\cal DL}o$) distribution is given by
\begin{equation}\label{Kn}
K_n=\sqrt{n}\stackbin[\mu\leq k\leq M]{}{\max}\; \lvert F_n(k)-F_n(k,\hat{\vartheta}_n)\rvert,
\end{equation}
where
\begin{equation}\label{edf}
F_n(k)=n^{-1}\sum_{i=1}^{n}1\{x_i\leq k\}
\end{equation}
is the EDF of the sample $x_1,\dots,x_n$, of size $n$, in a sample value $k$; $\hat{\vartheta}_n=\hat{\vartheta}_n(x_1,\dots,x_n)$ is the suitable estimator (in this study, the maximum likelihood estimator obtained previously, as described in section \ref{est}) of the unknown parameter vector $\vartheta=(\alpha,\lambda,\mu)$; and $M$ is the sample maximum $$M=x_{max}=\stackbin[1\leq i\leq n]{}{\max}\;x_i.$$
The null hypothesis can be expressed again as
$$H_0: the\;data\;follow\;the\;discrete\;{\cal DGP}\;(or\;{\cal DL}o)\;model,$$
Then, the procedure is as follows \citep{Prieto2014}:
\begin{enumerate}[(1)]
\addtolength{\itemsep}{-3.5mm}
\item calculate the empirical $KS$ statistic for the observed data using Eq.(\ref{Kn});
\item generate by simulation, using Eq.(\ref{guantil}), enough synthetic data sets with the same sample size $n$ as the observed data (in this study, we generated 10000 datasets);
\item fit each synthetic dataset by maximum likelihood (see section \ref{est}), and obtained its theoretical cumulative distribution function cdf using Eq.(\ref{edf});
\item calculate the $KS$ statistic for each synthetic dataset, using its own theoretical cdf (instead of the EDF) and the synthetic dataset (instead of the observed data) in Eq.(\ref{Kn});
\item calculate the $p$-value as the fraction of synthetic datasets with a $KS$ statistic greater than the empirical $KS$ statistic;
\item null hypothesis can be rejected at the 0.05 level of significance if the $p$-value calculated is less than 0.05.
\end{enumerate}

\section{Results}\label{results}
Table \ref{par} shows the parameter estimates $\hat{\alpha}$ and $\hat{\lambda}$, and their standard errors: (1) from the ${\cal DGP}$
distribution, fitted to the first variable data analyzed (number of accidents on blackspots), with $\hat{\mu}=x_{min}=3$; and (2) from the ${\cal DL}o$ distribution, 
fitted to the second variable data, (number of deaths on blackspots); both models fitted by maximum likelihood, in the period 2003-2007. It can be seen that most of the estimates are statistically significant at a 0.05 level of significance, assuming the asymptotic normality of the maximum likelihood estimates.\\
Table \ref{chi} shows the values of the chi-square statistic, the degree of freedom and the corresponding chi-square critical values, from ${\cal DGP}$ and ${\cal DL}o$ distributions fitted, respectively to variable 1 and variable 2 data, in 2003-2007. It can be seen that the values of chi-square statistic are less than the corresponding chi-square critical values, except in the case of variable 1 in the year 2003. In addition, table \ref{chi} shows the one tailed (right-tail) probability value ($p$-value) for the chi-square test, corresponding to both variables analyzed in each year from 2003 to 2007, confirming the results obtained: all the $p$-values are greater than 0.05, except in the case of number of accidents (variable 1) in the year 2003. It means that ${\cal DGP}$ model can not be ruled out with the 0.05 level of significance in four of the five years considered and can be ruled out with that level of significance in 2003. It also means that ${\cal DL}o$ model can not be ruled out with the 0.05 level of significance in all the years considered.\\
Table \ref{KS} shows the values of the empirical $KS$ statistic
and the $p$-values obtained by bootstrap resampling. It can be seen that all the $p$-values obtained are greater
than 0.05, therefore the discrete $KS$ GOF test indicates that our null hypothesis ($H_0$: the data follow the discrete ${\cal DGP}$,
or ${\cal DL}o$, model) can not be rejected at a 0.05 level of significance, in both variables (number of accidents and number of deaths), in the period 2003-2007.\\
In summary, we have checked the maximum likelihood fit of the discrete generalized Pareto model (${\cal DGP}$) to the number of road accidents on Spanish blackspots for each year of the period 2003-2007. Also, we have checked the maximum likelihood fit of discrete Lomax model (${\cal DL}o$) to the number of deaths on Spanish blackspots in the same period. In both cases, using the Chi-Square goodness-of-fit test and the discrete Kolmogorov-Smirnov goodness-of-fit test. The results indicate that those probabilistic models can be useful to describe the road accident blackspots datasets analyzed.
\begin{table}[p]\tiny
  \renewcommand{\tablename}{\footnotesize{Table}}
  \renewcommand\arraystretch{1.4}
  \setlength{\tabcolsep}{5 mm}
  \caption{\label{par}Parameter estimates from ${\cal DGP}$ and ${\cal DL}o$ models to the number of accidents (variable 1, $\mu=3$) and deaths (variable 2), respectively, on Spanish blackpots dataset, by maximum likelihood (standard error in parenthesis).}
  \centering
        \begin{tabular*}{1.0\textwidth}{l c c c c c c}
        \\[-1ex]
     \hline
                         &                        &  {\bf 2003}     & {\bf 2004}  &  {\bf 2005}      &  {\bf 2006}     &  {\bf 2007}  \\
     \hline
                         &{\bf $\hat{\alpha}$  }  &   3.8227        &   3.2601    &   3.3883         &   4.0439        &   3.5710        \\
${\cal DGP}$ model       &                        &   (0.6398)      &   (0.5140)  &   (0.5443)       &   (0.7178)      &   (0.6093)      \\
(no. accidents)          &{\bf $\hat{\lambda}$}   &   0.2295        &   0.2933    &   0.2719         &   0.2182        &   0.2547        \\
                         &                        &   (0.0482)      &   (0.0599)  &   (0.0559)       &   (0.0479)      &   (0.0552)      \\[0.5ex]
     \hline
                         &{\bf $\hat{\alpha}$  }  &   6.5547        &   13.8596   &   5.4875         &   4.3400        &   10.8251       \\
${\cal DL}o$ model                &                        &   (2.0654)      &   (9.8951)  &   (1.6803)       &   (1.1572)      &   (5.8841)      \\
(no. deaths)             &{\bf $\hat{\lambda}$ }  &   0.3142        &   0.1285    &   0.3811         &   0.5355        &   0.2039        \\
                         &                        &   (0.1181)      &   (0.0999)  &   (0.1435)       &   (0.1857)      &   (0.1245)      \\[0.5ex]
     \hline
     \end{tabular*}
\end{table}
\begin{table}[p]\tiny
  \renewcommand{\tablename}{\footnotesize{Table}}
  \renewcommand\arraystretch{1.4}
  \setlength{\tabcolsep}{4.6 mm}
  \caption{\label{chi}Chi-square statistic values and the corresponding $p$-values, from ${\cal DGP}$ and ${\cal DL}o$ models, to the number of accidents and deaths on Spanish blackpots respectively, from 2003 to 2007. $\chi^2>\chi_{0.95,k-r-1}^{2}$ or $p$-value$<0.05$ indicates than the model can be ruled out with the 0.05 level of significance.}
  \centering
        \begin{tabular*}{1.0\textwidth}{l c c c c c c}
        \\[-1ex]
     \hline
                   &                                       &  {\bf 2003}     & {\bf 2004}  &  {\bf 2005}      &  {\bf 2006}     &  {\bf 2007}    \\
     \hline
${\cal DGP}$ model          &{\bf $\chi^2$                }         &   17.930        &   2.608     &   5.537          &   10.397        &   4.903        \\
(no. accidents)    &{\bf $df=k-r-1$              }         &   6             &   6         &   5              &   5             &   6            \\
                   &{\bf $\chi_{0.95,k-r-1}^{2}$ }         &   12.592        &   12.592    &   11.071         &   11.071        &   12.592        \\[1.0ex]
                   &$p$-value                              &   0.0064        &   0.8561    &   0.3539         &   0.0647        &   0.5563        \\[0.5ex]
     \hline
${\cal DL}o$ model          &{\bf $\chi^2$                }         &   3.639         &   0.590     &   0.556          &   0.203         &   0.918        \\
(no. deaths)       &{\bf $df=k-r-1$              }         &   1             &   1         &   1              &   1             &   1            \\
                   &{\bf $\chi_{0.95,k-r-1}^{2}$ }         &   3.841         &   3.841     &   3.841          &   3.841         &   3.841         \\[1.0ex]
                   &$p$-value                              &   0.0564        &   0.4425    &   0.4560        &   0.6527         &   0.3380        \\[0.5ex]
     \hline
     \end{tabular*}
\end{table}
\begin{table}[p]\tiny
  \renewcommand{\tablename}{\footnotesize{Table}}
  \renewcommand\arraystretch{1.4}
  \setlength{\tabcolsep}{5.2 mm}
  \caption{\label{KS}Empirical $KS$ statistics and bootstrap $p$-values corresponding to both variables analyzed in each year from 2003 to 2007. $p$-value$<0.05$ indicates than the model can be ruled out with the 0.05 level of significance.}
  \centering
        \begin{tabular*}{1.0\textwidth}{l c c c c c c}
        \\[-1ex]
     \hline
                   &                         &  {\bf 2003}     & {\bf 2004}  &  {\bf 2005}      &  {\bf 2006}     &  {\bf 2007}    \\
     \hline
${\cal DGP}$ model          &         {\bf $KS$}      &   0.3088        &   0.1712    &   0.3950         &   0.4810        &   0.1867       \\
(no. accidents)    &{\bf $p$-value }         &   0.3322        &   0.8087    &   0.1351         &   0.0518        &   0.7640        \\[0.5ex]
     \hline
${\cal DL}o$ model          &{\bf $KS$      }         &   0.1361        &   0.1152    &   0.0824         &   0.0475        &   0.0978       \\
(no. deaths)       &{\bf $p$-value }         &   0.2606        &   0.3987    &   0.6226         &   0.9047        &   0.2962        \\[0.5ex]
     \hline
     \end{tabular*}
\end{table}
\section{Conclusions}\label{conclusions}

We found a three parameter probability distribution that we can use to model the number of road accidents on blackspots: the discrete generalized Pareto distribution.

We found a two parameter probability distribution that we can also use to model road traffic networks events, in particular the number of deaths on road accident blackspots: the discrete Lomax distribution.

We considered road accidents on Spanish blackspots data, from Spanish General Directorate of Traffic (DGT), where blackspots are identificated as road sections of 100 meters with three or more traffic accidents in a one-year period, and where a death is included if ocurred within 30 days after the accident. We modelled 16552 accidents, ocurred on Spanish road blackspots in the period 2003-2007; analyzed two discrete variables: the number of accidents and the number of deaths; fitted the discrete generalized Pareto and the discrete Lomax distributions, respectively to the data, by maximum likelihood; and tested the goodness-of-fit of those models by a Chi-square test and by a Kolmogorov-Smirnov test method based on bootstrap resampling.

In this study, we found that road traffic networks events, specifically road accident blackspots events, can be described by simple probabilistic models: the discrete generalized Pareto and the discrete Lomax distributions.
\section*{Acknowledgements}

The authors thank to Ministerio de Econom\'{\i}a y Competitividad
(projects ECO2010-15455 (FP and JS) and ECO2009-14152 (EGD)) for partial support of
this work. 

\bibliographystyle{plainnat}

\begin{thebibliography}{99}

\bibitem[Aguero-Valverde(2013)]{AgueroValverde2013}
Aguero-Valverde, J., 2013. Full Bayes Poisson gamma, Poisson lognormal, and zero inflated random effects models: comparing the precision of crash frequency estimates. Accident Analysis and Prevention 50, 289-297.
\bibitem[Alemany et al.(2013)]{Alemany2013}
Alemany, R., Ayuso, M., Guill\'en, M., 2013. Impact of road traffic injuries on disability rates and long-term care costs in Spain. Accident Analysis and Prevention 60, 95-102.
\bibitem[Arnold(1983)]{Arnold1983}
Arnold, B.C., 1983. Pareto distributions. International Co-operative Publishing House, Fairland, Maryland.
\bibitem[Asadi et al.(2001)]{Asadi2001}
Asadi, M., Rao, C.R., Shanbhag, D.N., 2001. Some unified characterization results on generalized Pareto distributions. Journal of Statistical Planning and Inference 93, 29-50.
\bibitem[Brijs et al.(2007)]{Brijs2007}
Brijs, T., Karlis, D., Van den Boussche and Geert Wets, F., 2007. A Bayesian model for ranking hazardous road sites. Journal of the Royal Statistical Society: Series A 170(4), 1001-1017.
\bibitem[Cafiso and Di Silvestro(2011)]{Cafiso2011}
Cafiso, S., Di Silvestro, G., 2011. Evaluation of safety identification methods for low-volume roads using Monte Carlo simulation.
Journal Transportation Research Record: Journal of the Transportation Research Board 2203, 106-115.
\bibitem[Chen et al.(2011)]{Chen2011}
Chen, Y., Liu, C., Wu, H., un, W., 2011. Identification of black spot on traffic accidents and its spatial associatoin analysis based on Geographic Information System. Seventh International Conference on Natural Computation, 143-150.
\bibitem[Cheng and Washington(2005)]{Cheng2005}
Cheng, W., Washington, S.P., 2005. Experimental evaluation of hotspot identification methods. Accident Analysis and Prevention 37, 870-881.
\bibitem[Deublein et al.(2013)]{Deublein2013}
Deublein, M., Schubert, M., Adey, B.T., Köhler, J., Faber, M.H., 2013. Prediction of road accidents: A Bayesian hierarchical approach. 
Accident Analysis and Prevention 51, 274-291.
\bibitem[DGT(2013a)]{DGT1}
D.G.T. 2013a. Direcci\'on General de Tr\'afico, Ministerio del Interior, Gobierno de Espa\~na. Estudios e informes de seguridad vial. Puntos negros 2003-2007.
\href{http://www.dgt.es/portal/es/informacion_carreteras/puntos_negros/anyos_anteriores/}{http://www.dgt.es/portal/es/informacion\_carreteras/ [28/10/2013].}
\bibitem[DGT(2013b)]{DGT2}
D.G.T., 2013b. Direcci\'on General de Tr\'afico, Ministerio del Interior, Gobierno de Espa\~na. Estad\'{\i}sticas e indicadores.
\href{http://www.dgt.es/portal/es/seguridad_vial/estadistica/}{http://www.dgt.es/portal/es/seguridad\_vial/estadistica/ [28/10/2013].}
\bibitem[Ekheden and Hössjer(2012)]{Ekheden2012}
Ekheden, E., Hössjer, O., 2012. Pricing catastrophe risk in life (re)insurance. Scandinavian Actuarial Journal, 1-16.
\bibitem[G\'omez-D\'eniz(2010)]{GomezDeniz2010}
G\'omez-D\'eniz, E., 2010. Another generalization of the geometric distribution. Test 19, 399-415.
\bibitem[G\'omez-D\'eniz and Calder\'{\i}n-Ojeda(2011)]{GomezDeniz2011}
G\'omez-D\'eniz, E.; Calder\'{\i}n-Ojeda, E., 2011. The discrete Lindley distribution: properties and applications.
Journal of Statistical Computation and Simulation 81(11), 1405-1416.
\bibitem[G\'omez-D\'eniz et al.(2011)]{GomezDeniz2011b}
Gómez-Déniz, E., Sarabia, J.M., Calderín-Ojeda, E., 2011. A new discrete distribution with actuarial applications. 
Insurance: Mathematics and Economics 48(3), 406-412.
\bibitem[Gregoriades and Mouskos(2013)]{Gregoriades2013}
Gregoriades, A., Mouskos, K.C., 2013. Black spots identification through a Bayesian Networks quantification of accident risk index. 
Transportation Research Part C 28, 28-43.
\bibitem[Hauer et al.(2002)]{Hauer2002}
Hauer, E., Harwood, D.W., Council, F.M., Griffith, M., 2002. Estimating safety by the empirical bayes method: a tutorial. 
Transportation Research Record: Journal of the Transportation Research Board 1784, 126–131.
\bibitem[Henze(1996)]{Henze1996}
Henze N., 1996. Empirical-distribution-function goodness-of-fit tests for discrete models. The Canadian Journal of Statistics 24(1), 81-93.
\bibitem[Hoque et al.(2007)]{Hoque2007}
Hoque, M.S., Moniruzzaman, S.M., Mahmud, S.M.S., 2007. Effectiveness of black spot treatments along Dhaka-Aricha highway. 
Journal of Civil Engineering 35(2), 93-104.
\bibitem[Johnson and Kotz(1982)]{JohnsonandKotz1982}
Johnson, N.L. and Kotz, S., 1982. Developments in discrete distribution, 1969-1980. International Statistical Review 50, 71-101.
\bibitem[Jurenoks et al.(2008)]{Jurenoks2008}
Jurenoks, V., Jansons, V., Didenko, K., 2008. Investigation of Accident Black Spots on Latvian Roads Using Scan Statistics Method. 
22nd EUROPEAN Conference on Modelling and Simulation ECMS 2008, 3-6. 
\bibitem[Karlis and Xekalaki(2005)]{KarlisandXekalaki2005} 
Karlis, D. and Xekalaki, E., 2005. Mixed Poisson distributions. International Statistical Review 73, 35-58.
\bibitem[Kemp(2008)]{Kemp2008}
Kemp A.W., 2008. The discrete half-normal distribution. In: Advances in Mathematical and Statistical Modeling. Birkh\"{a}user, Basel, 353-365.
\bibitem[Kim et al.(2007)]{Kim2007}
Kim, D-G., Lee, Y., Washinton, S., Choi, K., 2007. Modeling crash outcome probabilities at rural intersectoins: application of hierarchical binomial logistic models. Accident Analysis and Prevention 39, 125-134.
 \bibitem[Krishna and Singh(2009)]{Krishna2009}
Krishna, H., Singh, P.,2009. Discrete Burr and discrete Pareto
distributions. Statistical Methodology 6(2), 177-188.
\bibitem[Li(2012)]{Li2012} Li, S., 2012. 
Traffic safety and vehicle choice: quantifying the effects of the 'arms race' on American roads. Journal of Applied Econometrics 27, 34-62.
\bibitem[Mandloi and Gupta(2003)]{Mandloi2003}
Mandloi, D., Gupta, R., 2003. Evaluation of accident black spots on roads using Geographical Information Systems (GIS).
6th Annual International Conference, Map India, Paper no. 38.
\bibitem[Mat\'{\i}rnez et al.(2013)]{Matirnez2013}
Mat\'{\i}rnez A., Mántaras, D.A., Luque, P. 2013. Reducing posted speed and perceptual countermeasures to improve safety in road stretches with a high concentration of accidents. Safety Science 60, 160-168.
\bibitem[Meuleners et al.(2008)]{Meuleners2008}
Meuleners, L.B., Hendrie, D., Lee, A,H., Legge, M., 2008. Effectiveness of the black spot programs in Western Australia.
Accident Analysis and Prevention 40, 1211-1216.
\bibitem[Nakagawa and Osaki(1975)]{Nakagawa1975}
Nakagawa, T., Osaki, S., 1975. The discrete Weibull distribution. IEEE Transactions of Reliability 24(5), 300-301.
\bibitem[Pei and Ding(2005)]{Pei2005}
Pei, J., Ding, J., 2005. Improvement in the Quality Control Method to Distinguish the Black Spots of the Road,
Proceedings of the Eastern Asia Society for Transportation Studies 5, 2106 - 2113.
\bibitem[Pearson(1900)]{Pearson1900}
Pearson, k., 1900. On the criterion that a given system of deviations from the probable in the case of a correlated system of variables is such that it can be
reasonably supposed to have arisen from random sampling. Philosophical Magazine Series 5, 50(132),157-175.
\bibitem[Prieto et al.(2014)]{Prieto2014}
Prieto. F., Sarabia, J.M., Sáez, A.J., 2014. Modelling major failures in power grids in the whole range.
International Journal of Electrical Power \& Energy Systems 54, 10-16.
\bibitem[Roy(2004)]{Roy2004}
Roy, D., 2004. Discrete Rayleigh distribution. IEEE Transactions of Reliability 53(2), 255-260.
\bibitem[Sabel et al.(2005)]{Sabel2005}
Sabel, C., Kingham, S., Nicholson, A., Bartie, P., 2005. Road Traffic Accident Simulation Modelling - A Kernel Estimation Approach. The 17th Annual Colloquium of the Spatial Information Research Centre University of Otago. Dunedin, New Zealand, 67–75.
\bibitem[Warde and Katti(1971)]{WardeandKatti1971}
Warde, W.D. and Katti, S.K., 1971. Infinite divisibility of discrete distributions II. The Annals of Mathematical Statistics 42(3), 1088-1090.
\bibitem[WHO(2013)]{WHO}
W.H.O. World Health Organization.
\href{http://who.int/mediacentre/factsheets/fs310/en/}{http://www.who.int/}
\bibitem[Mathematica (version 8.0)]{Mathematica2010}
Wolfram Research, Inc., 2010. Mathematica. Version 8.0, Champaign, IL. 
\bibitem[Xekalaki(1983)]{Xekalaki1983}
Xekalaki, E., 1983. Hazard functions and life distributions in discrete time. 
Communications in Statistics-Theory and Methods 12(21), 2503-2509.

\end{thebibliography}

\end{document}